\documentclass[a4paper]{article}

\usepackage{spie}
\usepackage{graphicx}
\usepackage{aasabrev}

\newcommand{\micron}{\mbox{$\,\mu\mathrm{m}$}}

\title{The SPIFFI image slicer: Revival of image slicing with plane mirrors}

\author{Matthias Tecza, Niranjan Thatte, Frank Eisenhauer, Sabine
Mengel, Claudia R\"ohrle and\\Klaus Bickert
\skiplinehalf
Max-Planck-Institut f\"ur extraterrestrische Physik,\\
Postfach 1603, D-85740 Garching, Germany}

\authorinfo{Further author information: Send correspondence to M.\ Tecza,
e-Mail:~tecza@mpe.mpg.de}

\begin{document}

\maketitle

\begin{abstract}
SPIFFI (SPectrometer for Infrared Faint Field Imaging) is the integral
field spectrograph of the VLT-instrument SINFONI (SINgle Far Object
Near-infrared Investigation).  SINFONI is the combination of SPIFFI
with the ESO adaptive optics system MACAO (Multiple Application Concept
for Adaptive Optics) offering for the first time adaptive optics
assisted near infrared integral field spectroscopy at an
$8\,\mathrm{m}$-telescope.  SPIFFI works in the wavelength ranger from
$1.1$ to $2.5\micron$ with a spectral resolving power ranging from
$R=2000$ to $4500$.  Pixel scale ranges from $0.25$ to $0.025$ seconds
of arc.  The SPIFFI field-of-view consists of $32\times32$ pixels
which are rearranged with an image slicer to a form a long slit.

Based on the 3D slicer concept with plane mirrors, an enhanced image
slicer was developed.  The SPIFFI image slicer consists of two sets of
mirrors, called the \emph{small} and the \emph{large} slicer.  The
small slicer cuts a square field of view into 32 slitlets, each of
which is 32 pixels long.  The large slicer rearranges the 32 slitlets
into a 1024 pixels long slit.  The modifications to the 3D slicer
concept affect the angles of the plane mirrors of small and large
slicer and lead to an improved slit geometry with very little light
losses.  At a mirror width of $0.3\,\mathrm{mm}$ the light loss is
$<5\%$.  All reflective surfaces are flat and can be manufactured with
a high surface quality.  This is especially important for the adaptive
optics mode of SINFONI.  We explain the concept of the SPIFFI mirror
slicer and describe details of the manufacturing process.
\end{abstract}


\section{INTRODUCTION}\label{Sec:Introduction}
Classical imaging spectroscopy uses scanning techniques to obtain both
spatial and spectral information of a two-dimensional field-of-view on
the sky.  Either wavelength-scanning with a Fabry-Perot camera or
slit-scanning with a long-slit spectrograph are used to generate a
data cube.  Fabry-Perot imaging covers a large field-of-view with a
limited wavelength range while long-slit spectroscopy yields a
spectrum over a wide wavelength range and a limited field-of-view.
Depending on the object of interest and scientific proposal either
scanning technique might be well suited.  However, if compact objects
are observed, only a fraction of the data taken by scanning techniques
contains useful information because most of the large field-of-view or
long slit length covers sky.

For compact objects the technique of integral field spectroscopy is
better suited.  With an integral field spectrograph the spectra of all
pixels in a two-dimensional field can be obtained simultaneously
therefore reducing the integration time over scanning techniques.  But
this multiplex advantage is not the only advantage of integral field
spectroscopy.  Especially in the near-infrared wavelength range the
variability of the atmosphere makes it difficult to generate a data
cube from data which were taken in a time sequence.  All spectra taken
with an integral field spectrograph are affected equally by the
atmosphere resulting in a much improved data quality.

Presently three major techniques are used for integral field units.
\begin{itemize}
\item The TIGER-concept \cite{TIGER} uses a microlens array to dissect
the field-of-view into individual pixels.  The microlens array has a
very high filling factor of almost $100\%$.  Each microlens creates an
image of the telescope pupil.  Because of their small size they are
called \emph{micro-pupils}.  All micro-pupils form an array and
because of their small size the filling factor of the micro-pupil
array is very small.  The idea of the TIGER concept is to image the
micro-pupil array onto the array detector but disperse the light of
each micro-pupil onto the unfilled portions of the detector.  This
concept has been successfully used in the visible wavelength range.
\item Another technique uses a fiber bundle to transform a
two-dimensional field into a slit which is fed into a long-slit
spectrograph.  To transform the field-of-view the fiber bundle has a
two-dimensional cross-section on one end and a one-dimensional
cross-section on the other end.  To increase the coupling efficiency
often a microlens array is used at the entrance end.  Examples for
instruments using this technique are INTEGRAL \cite{INTEGRAL} in the
visible wavelength range, and SMIRFS-II \cite{SMIRFS-II} and COHSI
\cite{COHSI} in the near infrared wavelength range.  Also the first
design of SPIFFI followed this approach \cite{SPIFFI-I}.
\item The third concept uses a mirror slicer which dissects the
field-of-view into slitlets and rearranges these slitlets to form a
slit which is fed into a long-slit spectrograph.  This concept was
first used in the integral field spectrograph 3D \cite{3D} which was
designed and built at the Max-Planck-Institut f\"ur extraterrestrische
Physik (MPE).  The new near-infrared integral field spectrograph
SPIFFI \cite{SPIFFI-II} of MPE uses this concept which is described
in this article.
\end{itemize}

Integral field spectroscopy became an interesting and useful technique
only with the advent of large size array detectors.  Only large array
detectors permit to do integral field spectroscopy with large
wavelength range in a sufficiently large field-of-view.

3D for example uses a NICMOS~3 detector \cite{NICMOS-3} with
$256\times256$ pixels of which $256$ rows correspond to the spectra of
256 spatial pixels in a $16\times16$ pixels large field-of-view.  With
the new Rockwell HAWAII detector \cite{HAWAII} with $1024\times1024$
pixels a four times bigger field-of-view can be covered and either the
spectral wavelength range or the spectral resolving power can be
quadruplicated.  To make use of these new detectors at MPE the new
instrument SPIFFI using a HAWAII array was designed and is currently
under construction.  Its scientific applications range from
investigations of our solar systems to observations of cosmologically
distant galaxies \cite{SPIFFI-science}.  In this article we describe
the design and fabrication of the SPIFFI image slicer, the ``heart''
of SPIFFI.

\section{3D IMAGE SLICER}\label{Sec:3D}
3D, designed and built by MPE, is the first near-infrared integral
field spectrograph which utilizes a new concept for the image slicer.
The 3D image slicer uses a system of plane mirrors to transform a
two-dimensional field-of-view into a pseudo-slit.  A NICMOS~3 detector
is used in such a way that spectra of 256 pixels in a $16\times16$
pixels large field-of-view can be obtained simultaneously.

\begin{figure}[ht]
\begin{center}
\includegraphics[width=0.8\textwidth]{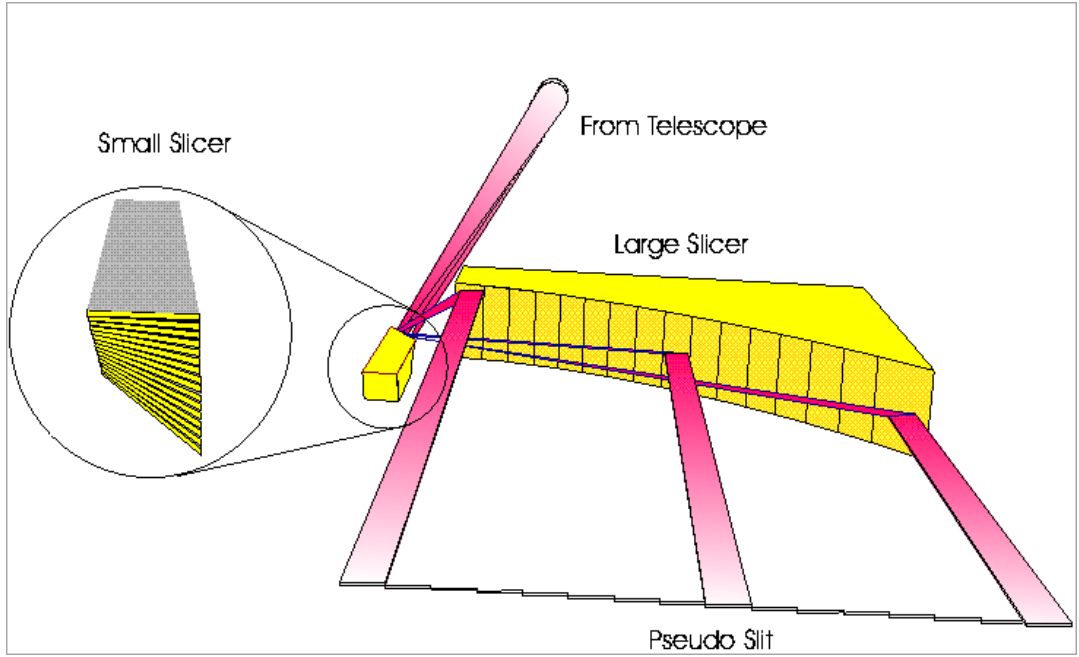}
\end{center}
\caption{\label{Fig:3D-slicer}
A perspective view of the 3D-slicer illustrating the image slicer
concept.  The small slicer is placed in a focal plane and dissects the
square field-of-view into 16 slitlets, each sixteen pixels long.  The
large slicer reflects the ray fan coming from the small slicer to form
a pseudo-slit.  The small and large slicer consist of 16 mirrors each,
on the left is a blow-up of the small slicer showing the 16 mirror
facets with different tilt-angles.}
\end{figure}

The 3D image slicer consists of two sets of mirrors.  The first set of
mirrors, the \emph{small slicer}, is a stack of 16 mirrors placed in
the focal plane of the telescope.  Th mirrors are only
$0.4\,\mathrm{mm}$ wide and each mirror is tilted by a different angle
but all the same direction (see Figure \ref{Fig:3D-slicer}).  The 16
different tilt-angles lead to a sliced field-of-view consisting of 16
stripes fanned out over $>90^\circ$.  The projected length of the 16
stripes in the telescope focal plane is $6.4\,\mathrm{mm}$,
corresponding to $16$ pixels of $0.4\,\mathrm{mm}$ width.

After the small slicer the fan of 16 stripes hits the second set of
mirrors, called the \emph{large slicer}.  The large slicer reflects
the rays from to form a pseudo-slit.  The tilt-angles of the 16
mirrors of the small and large slicer are such that the position of
the telescope pupil is conserved.  To achieve this the centers of the
large slicer mirrors also have lie on a hyperbola.  The large slicer
mirrors are placed at a distance from the small slicer where the 16
stripes don't overlap anymore, resulting in a contiguous pseudo-slit of
16 slitlets arranged in a staircase pattern (see Figure
\ref{Fig:3D-slicer}).  The total length of the pseudo-slit is
$16\times16\times0.4\,\mathrm{mm}=102.4\,\mathrm{mm}$.

Neither the small slicer mirrors nor the large slicer mirrors are
located entirely in the telescope focal plane.  Defocusing at the
small slicer and especially at the large slicer leads to a not perfect
slicing efficiency of the 3D image slicer.  At the small slicer, due
to the tilt-angle, the defocus spot outside the mirror center is
larger than the mirror width of $0.4\,\mathrm{mm}$.  At the large
slicer, due to the distance between the small and large slicer, the
defocus spot extends over the mirror length of
$16\times0.4\,\mathrm{mm}$.  Both effects lead to light losses through
the image slicer.  The slicing efficiency depends on the tilt-angle
of the small slicer mirrors and the distance between the small and
large slicer.  Because a smaller tilt-angle at the small slicer leads
to an increase of the distance between the small and large slicer an
optimal combination of tilt-angle and distance exists.  The 3D slicer,
which is designed to give $0.5\,\mathrm{arcsec}$ pixels at
$4\,\mathrm{m}$-class telescopes has an efficiency of $94\%$.

\section{SPIFFI IMAGE SLICER}\label{Sec:SPIFFI}
For the SPIFFI instrument a Rockwell HAWAII detector with
$1024\times1024$ detector elements is used.  With this detector size
an image slicer with $32\times32$ pixels can be realized.  Because the
3D slicer cannot be used for this format a new image slicer has to be
built for SPIFFI.  Also, the SPIFFI instrument is designed to be used
on an $8\,\mathrm{m}$-telescope which puts additional constraints on
the image slicer.  A simple scaling of the 3D slicer to 1024 spatial
pixels would result in a $409.6\,\mathrm{mm}$ long slicer with an
efficiency of $\approx80\%$.  Both the size and the efficiency are not
acceptable for SPIFFI.

\subsection{Modified 3D Slicer}\label{Sec:SPIFFI-Concept}
\begin{figure}[t]
\begin{center}
\includegraphics[width=0.85\textwidth]{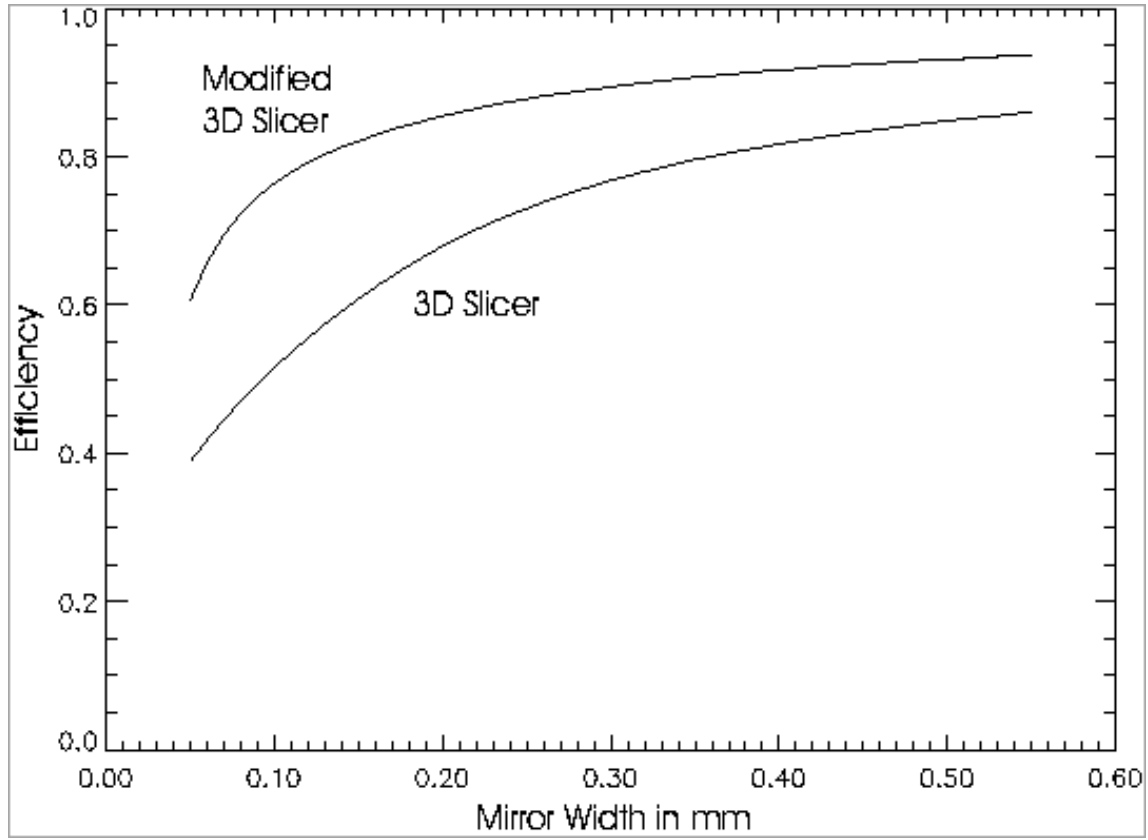}
\end{center}
\caption{\label{Fig:Losses}
Light losses of a 3D type slicer and a modified 3D type slicer for a
$32\times32$ pixels large field-of-view.  Pixel size on the sky is
0.25 seconds of arc, telescope diameter is $8\,\mathrm{m}$.}
\end{figure}

The only way to reduce the size of the SPIFFI image slicer is to
decrease the pixel size, i.e.\ the mirror width of the small slicer
mirrors.  However, decreasing the mirror width causes the efficiency
of the slicer to decrease.  Two effects are adding to yield a rapid
decrease in slicer efficiency (see Figure \ref{Fig:Losses}).  The
fraction of the defocus spot area of a pixel area increases with
decreasing mirror width.  In addition if the pixel size on the sky has
to stay constant, the beam in the telescope focal plane has a larger
f-ratio, increasing the defocus spot diameter even more.  Another
limitation is the feasibility of mirrors with a width
$<0.2\,\mathrm{mm}$.  For the SPIFFI image slicer a pixel size of
$0.3\,\mathrm{mm}$ is chosen, resulting in a pseudo-slit length of
$307.2\,\mathrm{mm}$.  This is a compromise between a still feasible
mirror width and the overall size of the slicer.

\begin{figure}[t]
\begin{center}
\includegraphics[width=0.85\textwidth]{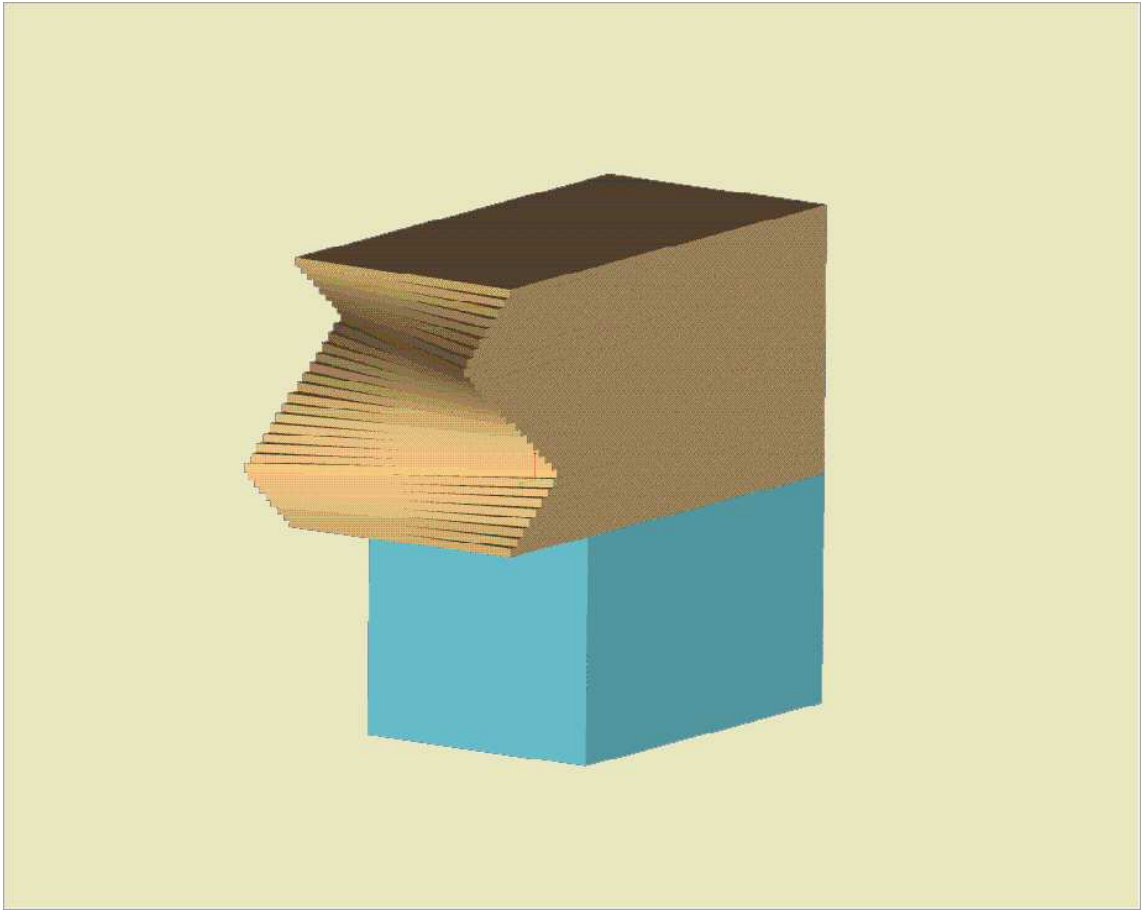}
\end{center}
\caption{\label{Fig:SPIFFI-smallslicer}
The \emph{small slicer} of the SPIFFI image slicer with 32 mirrors
facets.  The 32 mirrors are made from Zerodur using classical optical
polishing techniques.  The are held together only through optical
contacting.  The stack of mirrors is also optically contacted to the
Zerodur base of the small slicer.}
\end{figure}

\begin{figure}[t]
\begin{center}
\includegraphics[width=0.85\textwidth]{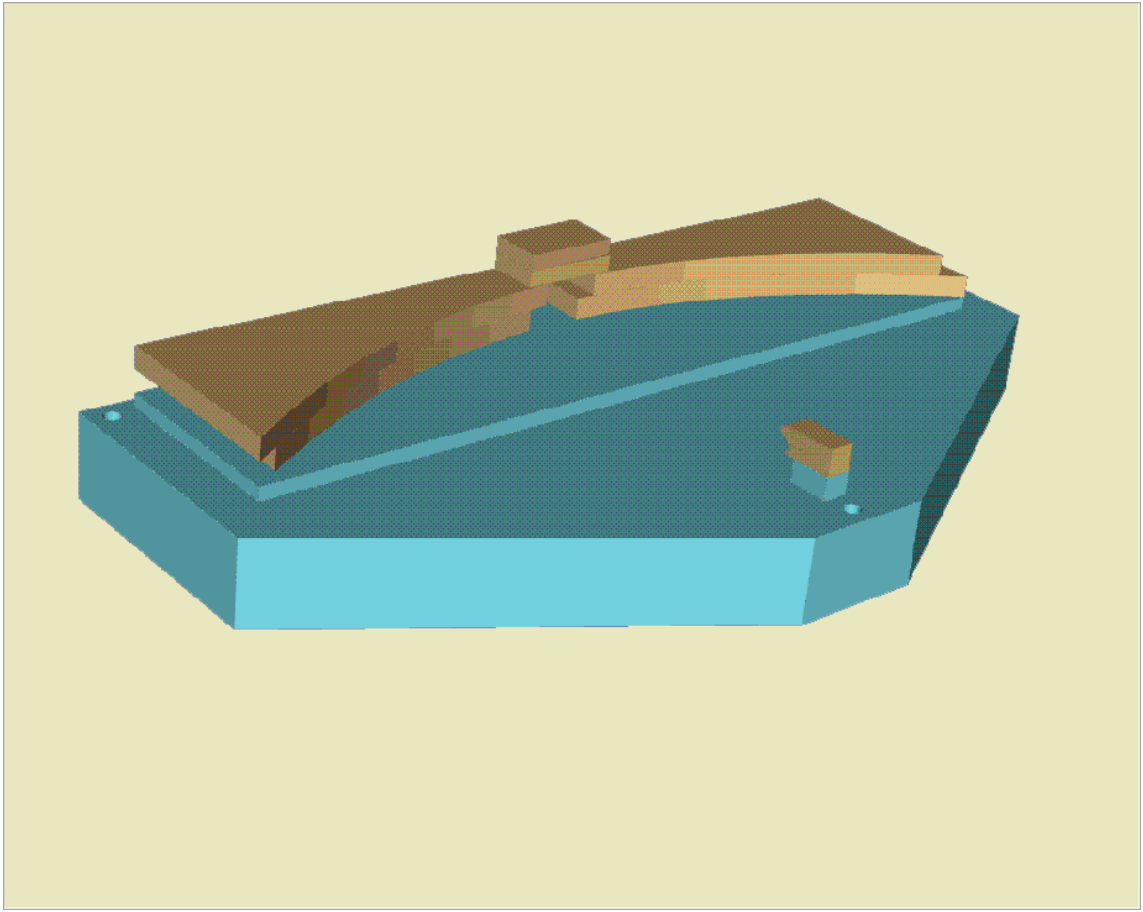}
\end{center}
\caption{\label{Fig:SPIFFI-slicer}
The SPIFFI image slicer.  Both small and large slicer are made
entirely from Zerodur.  Also the base plate and the spacer plates are
made from Zerodur.  No glue is used in the assembly of the image
slicer.  All mirrors are optically contacted.}
\end{figure}

However, as can be seen from Figure \ref{Fig:Losses} the efficiency of
a 3D-type slicer with $32\times32$ pixels and $0.3\,\mathrm{mm}$
mirror width is only $\approx80\%$.  To improve the slicer efficiency
three major changes to the 3D slicer concept have been applied (see
also Figures \ref{Fig:SPIFFI-smallslicer} and
\ref{Fig:SPIFFI-slicer}):
\begin{enumerate}
\item symmetric slicer layout,
\item additional mirror tip-angle orthogonal to the tilt-angle of the
3D slicer,
\item maximum tilt-angle of $32^\circ$ at small slicer.
\end{enumerate}
The 3D-slicer has an asymmetric layout (see Figure
\ref{Fig:3D-slicer}).  If this layout was used in the SPIFFI slicer,
the maximal distance between small and large slicer would equal the
pseudo-slit length of $307.2\,\mathrm{mm}$.  The symmetric layout
reduces the maximum distance between the small slicer and the large
slicer by a factor of 2 to only half the pseudo-slit length, thus
reducing the maximal defocus spot diameter by a factor of 2.

Because the 3D-slicer uses only one tilt-angle the pseudo-slit has a
staircase-pattern where the corners of adjacent slitlets touch each
other (see also Figure \ref{Fig:3D-slicer}).  Also the mirror facets
of the large slicer intersect at this position.  Because these facets
are out of focus a fraction of the light from pixels at the end of
each slitlet is reflected by the wrong facet and is lost.  With an
additional tip-angle of the slicer mirrors orthogonal to the
tilt-angle it is possible to change the staircase pattern of the
pseudo-slit in 3D to an almost arbitrary slit pattern.  For SPIFFI we
chose a two-layer brick-wall pattern.  With this pattern the individual
slitlets of the pseudo slit don't touch each other anymore, and the
large slicer mirrors can be oversized to reflect the entire defocused
slitlet.

The modification explained in the above paragraph is only possible up
to a maximum distance between the small and large slicer.  If the
distance was to large the defocus spot diameter would be larger than
the maximal possible mirror size of the large slicer.  With a
two-layer brick-wall pattern the minimal tilt-angle has to be
$16^\circ$ and the distance between the small and large slicer would
be $292\,\mathrm{mm}$.  Because this would be too large to fit into
the SPIFFI optical train a maximum tilt-angle of $32^\circ$ was
chosen.

The modifications to the 3D slicer concept lead to a very high
efficiency of the SPIFFI image slicer.  At a maximum tilt-angle of
$32^\circ$ the losses at the small slicer are reduced without
increasing the losses at the large slicer.  With a mirror width of
$0.3\,\mathrm{mm}$ all the light coming from the small slicer is
reflected by the large slicer.  In Figure \ref{Fig:Losses} we show the
light losses for an image slicer according to the 3D-slicer concept
and according to the modified 3D-slicer concept.  Assumed is a
field-of-view with $32\times32$ pixels with an input beam f-ration of
f/30, corresponding to a pixel size of $0.25$ seconds of arc at an
$8\,\mathrm{m}$-telescope.  Even for a mirror width of
$0.3\,\mathrm{mm}$ the light losses of the SPIFFI image slicer are
only $\approx10\%$.

Another advantage of the symmetric slicer layout is that instead of 32
different slicer mirrors for the small and large slicer only 16
different slicer mirrors, but two of each, have to be made.  This
reduces the cost of the SPIFFI image slicer.  However, with a
symmetric layout the direct view from the telescope along the optical
axis onto the small slicer is blocked by the central mirrors of the
large slicer.  Therefore, the two central mirrors of the big slicer
are moved and placed on top of the top-layer of the large slicer
leaving a hole for the view onto the small slicer along the optical
axis.  Because these two slitlets have larger tip-angles the shadowing
losses at the small slicer are also larger.  However, in the
two-dimensional field-of-view these slitlets correspond to the top and
bottom slitlets.  The correlation between the small and large slicer
mirrors is such that slicer-efficiency is largest in the field-center
of SPIFFI and is less at the field-edges.

To reduce thermal background radiation the image slicer is cooled down
$77\,\mathrm{K}$, the working temperature of SPIFFI instrument.  Based
on experience with 3D this increases the sensitivity of SPIFFI
especially in the wavelength range above $2\,\mathrm{\mu m}$.  

In Figure \ref{Fig:SPIFFI-smallslicer} we show the small slicer, in
Figure \ref{Fig:SPIFFI-slicer} we show a view of the whole
SPIFFI-slicer.

\subsection{Fabrication}\label{Sec:SPIFFI-Fabrication}
The entire SPIFFI image slicer, including the base plate, is made from
Zerodur.  Because of its low linear thermal expansion coefficient the
slicer design needs not be corrected for the slicer working
temperature of $77\,\mathrm{K}$.  

The small as well as the large slicer consist of individual mirrors
which are manufactured using conventional optical polishing
techniques.  The achieved surface quality is of order $\lambda/8$
which is necessary for the adaptive optics mode of SINFONI.

The good surface quality also allows to optically contact the mirrors
to build the mirror stacks.  We have performed tests under vacuum and
at low temperatures which show that the optically contacted Zerodur
surfaces don't separate.  The same technique is used to mount the
slicer mirror stacks to the base plate.  Hence, the SPIFFI image
slicer is a completely monolithic device.  It is not subject to
thermal stresses originating from differential thermal expansion
coefficients as it would be the case if glue was used in the assembly
of the individual mirrors.

To achieve a high accuracy on the tip- and tilt-angles a set of master
prisms with the correct angles was made.  They are large enough to be
accurately polished and measured.  Using these master prisms all
slicer mirrors are fabricated.  Angles are polished to better than
$10$ seconds of arc, length dimensions are known better than
$0.01\,\mathrm{mm}$.

An accurate assembly of the slicer mirrors to the small and large
slicer is achieved by checking for interference fringes between a
reference plate and the individual slicer mirrors.  This way the high
accuracy of the individual slicer mirrors is conserved during the
assembly of the small and large slicer.  The assembly of the small and
large slicer on the base plate is done with the help of a traveling
microscope.

After the small and large slicer are assembled the mirror surfaces are
coated with gold.  To protect the sensitive gold layer the mirrors are
coated with a thin layer of fused silica.

\subsection{Mounting}
The SPIFFI image slicer is mounted in a housing which acts as the
holder as well as the light baffle.  Special care has to be taken to
compensate for the differential thermal expansion coefficient of the
Zerodur slicer and the Aluminium holder.  No stress shall be applied
to the slicer while at the same time keeping its position stable
within $0.05\,\mathrm{mm}$.

In the design used the SPIFFI image slicer is held only at its base
plate.  The base plate is held between lapped Invar pads avoiding any
point contact between the metal holder and the slicer.  The Invar pads
touching the base plate are allowed a lateral movement of
$\approx1\,\mathrm{mm}$ with respect to the Aluminium holder.  This is
achieved through a ball bearing between the pads and the Aluminium
holder.

The position of the SPIFFI image slicer in the light-path orthogonal
to the optical axis is not critical.  The optical layout of SPIFFI
allows to compensate for offsets in these directions and the machining
tolerances are small enough to place the slicer accurately enough.
However, to be able to focus the slicer it is necessary to adjust the
position of the slicer along the optical axis.  This is done with a
micrometer screw which allows to defocus the SPIFFI image slicer by
$\pm3\,\mathrm{mm}$.

\bibliography{ms}
\bibliographystyle{spiebib}

\end{document}